\documentclass[%
 aip,
 amsmath,amssymb,
 reprint,%
]{revtex4-1}

\usepackage{graphicx}
\usepackage{dcolumn}
\usepackage{bm}
\usepackage{lineno}
\usepackage[utf8]{inputenc}
\usepackage[T1]{fontenc}
\usepackage{mathptmx}
\usepackage[final]{changes}
\renewcommand{\refname}{}

\begin{document}

\preprint{}

\title[Cough droplet dispersion behind a walking person]{Effects of space sizes on the dispersion of cough-generated droplets from a walking person}

\author{Zhaobin Li}
 
\author{Hongping Wang}%

\author{Xinlei Zhang}

\author{Ting Wu}

\author{Xiaolei Yang}
\email{xyang@imech.ac.cn}
 
\affiliation{ 
The State Key Laboratory of Nonlinear Mechanics, Institute of Mechanics, Chinese Academy of Sciences, Beijing 100190, China 
}%

\affiliation{ 
School of Engineering Sciences, University of Chinese Academy of Sciences, Beijing 100049, China 
}%

\date{\today}

\begin{abstract}
The dispersion of viral droplets plays a key role in the transmission of COVID-19. In this work, we analyze the dispersion of cough-generated droplets in the wake of a walking person for different space sizes. The air flow is simulated by solving the Reynolds-Averaged Navier-Stokes equations, and the droplets are modelled as passive Lagrangian particles. Simulation results show that the cloud of droplets locates around and below the waist height of the manikin after two seconds from coughing, which indicates that kids walking behind an infectious patient are exposed to higher transmission risk than adults. More importantly, two distinct droplet dispersion modes occupying significantly different contamination regions are discovered.  A slight change of  space size is found being able to trigger the transition of dispersion modes even though the flow patterns are still similar. This shows the importance of accurately simulating the air flow in predicting the dispersion of viral droplets and implies the necessity to set different safe-distancing guidelines for different environments. 
\end{abstract}

\maketitle

COVID-19 can be transmitted via respiratory droplets when a person is close to a patient who is coughing or sneezing \cite{who2020modesofTransmission}. The recommended safe social distance is two meters. However, it may not be enough for different conditions \cite{droplets,Blocken2020TowardsAE}, as the dispersion of viral droplets is significantly affected by the surrounding air flows, which depend on indoor and outdoor environments, the motion of the person, and other factors.  For this reason, accurately predicting the air flow and understanding its influence on dispersion of droplets are essential for predicting the spread range of the virus and thus to further provide guidelines for social-distancing  for different environments. Various efforts have been reported since the epidemic of COVID-19, including the study on the cough droplet dispersion by wind \cite{droplets},  the enhancement of droplet transmission distance by walking/running/cycling in the outdoor environment \cite{Blocken2020TowardsAE}, and the effectiveness of face mask on preventing the droplets transmission \cite{dbouk2020mask}, among others. 

In this work, we analyze the dispersion of cough droplets behind a walking person using Computational Fluid Dynamics (CFD), focusing on the influence of the indoor space constraints on the droplet dispersion by conducting simulations with different space sizes. The results reveal that the pattern of droplet dispersion can be significantly altered by only a slight change of the air flow, which demonstrates the importance of accurately predicting the air flow in predicting virus transmission for different environments. 

The numerical investigation consists of two-steps, \textit{i.e.}, 1) solve the flow around the walking person, 2) simulate the transient evolution of the cough droplets using the air flow obtained from step 1 by assuming that the one-way coupling is valid \textit{i.e.,} the air flow is not affected by the motion of the droplets. The air flow is simulated by solving the incompressible Reynolds-Averaged Navier-Stokes (RANS) equations, as follows:

\begin{align}
    \nabla \cdot \mathbf{u} &= 0, \\
    \mathbf{u} \cdot \left( \nabla \mathbf{u} \right) & = -\frac{\nabla p}{\rho} + \nu_\textrm{eff} \nabla^2 \mathbf{u},  
\end{align}
where $\mathbf{u}$ denotes the velocity vector, $p$ is the pressure, $\rho$ is the density of air and $\nu_\textrm{eff}$ is the effective viscosity, including both the molecular viscosity and the turbulent eddy viscosity computed with the $k-\omega$ SST turbulence model \cite{menter1994two}. The equations are solved with \texttt{simpleFoam} solver of \texttt{OpenFOAM} \cite{openFoam}. 

The respiratory droplets are modelled as passive particles with three translational degrees of freedom.  The rotation and evaporation of the droplets and interactions between droplets are excluded. The dispersion of the droplets is simulated using the Lagrangian method with their motion governed by the following equations: 

\begin{align}
\frac{\textrm{d} \mathbf{x}}{\textrm{d} t} &= \mathbf{v}_\textrm{c} + \mathbf{v}_\textrm{t},  \\
 m \frac{\textrm{d} \mathbf{v}_\textrm{c}}{\textrm{d} t} &= m \mathbf{g} + \mathbf{F},
\end{align}
where $\textbf{x}$ is the instantaneous position of the particle,  $\mathbf{v}_\textrm{c}$ is the computed particle velocity and $\mathbf{v}_\textrm{t}$ is the stochastic velocity due to turbulence, $m$ is the particle mass, $\mathbf{g}$ is the gravity, and $\mathbf{F}$ is the flow force acting on the particle, including the buoyancy and the drag computed with the air flow. The perturbation velocity $\mathbf{v}_\textrm{t}$ is computed with the stochastic dispersion model of Gosman \textit{et al.} \cite{gosman1983aspects}, where the fluctuation in direction $i$ is computed as
\begin{equation}
    v_\textrm{t}^i = \sigma \sqrt{\frac{2k}{3}},
\end{equation}
with $\sigma \sim N(0,1)$ following the standard normal distribution and $k$ is the turbulence kinematic energy obtained from the simulation of the air flow. The equations are solved with \newline \texttt{icoUncoupledKinematicParcelFoam} of \texttt{OpenFOAM}.

\begin{figure}
	\centering
		\includegraphics[width=0.35\textwidth	]{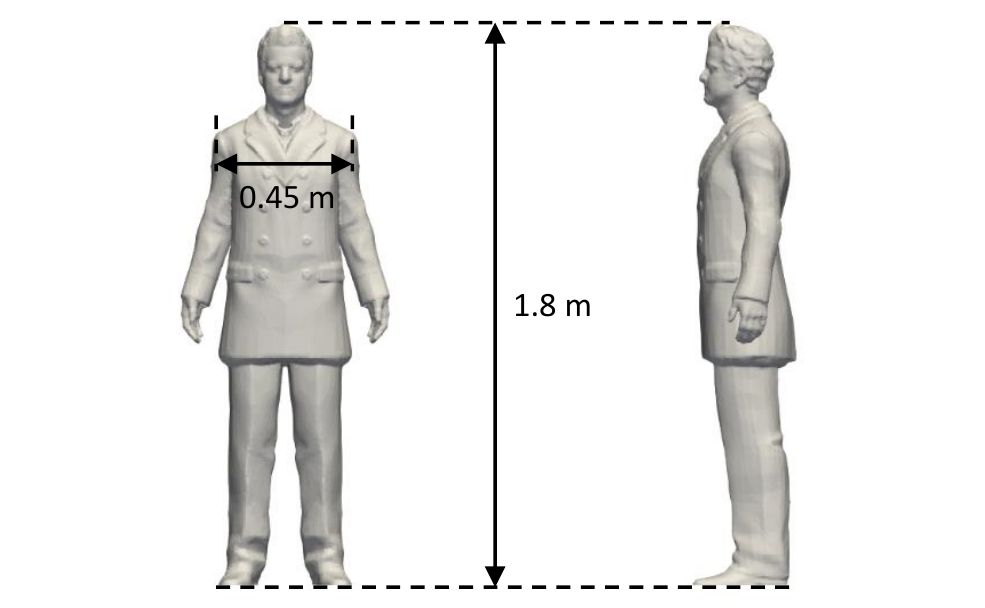}
	\caption{The manikin used in the simulations.}
	\label{fig:manikin}
\end{figure}

A human-shape manikin is employed in the simulation, which contains details of the human body and clothes to represents a medium-built male with the height of 1.8~m and the shoulder breadth of 0.45~m, as shown in Fig. \ref{fig:manikin}. To simplify the problem, the manikin is assumed as a rigid body without considering the motion of arms, legs, and other body parts relative to the overall movement.

\begin{figure}
	\centering
		\includegraphics[width=0.35\textwidth	]{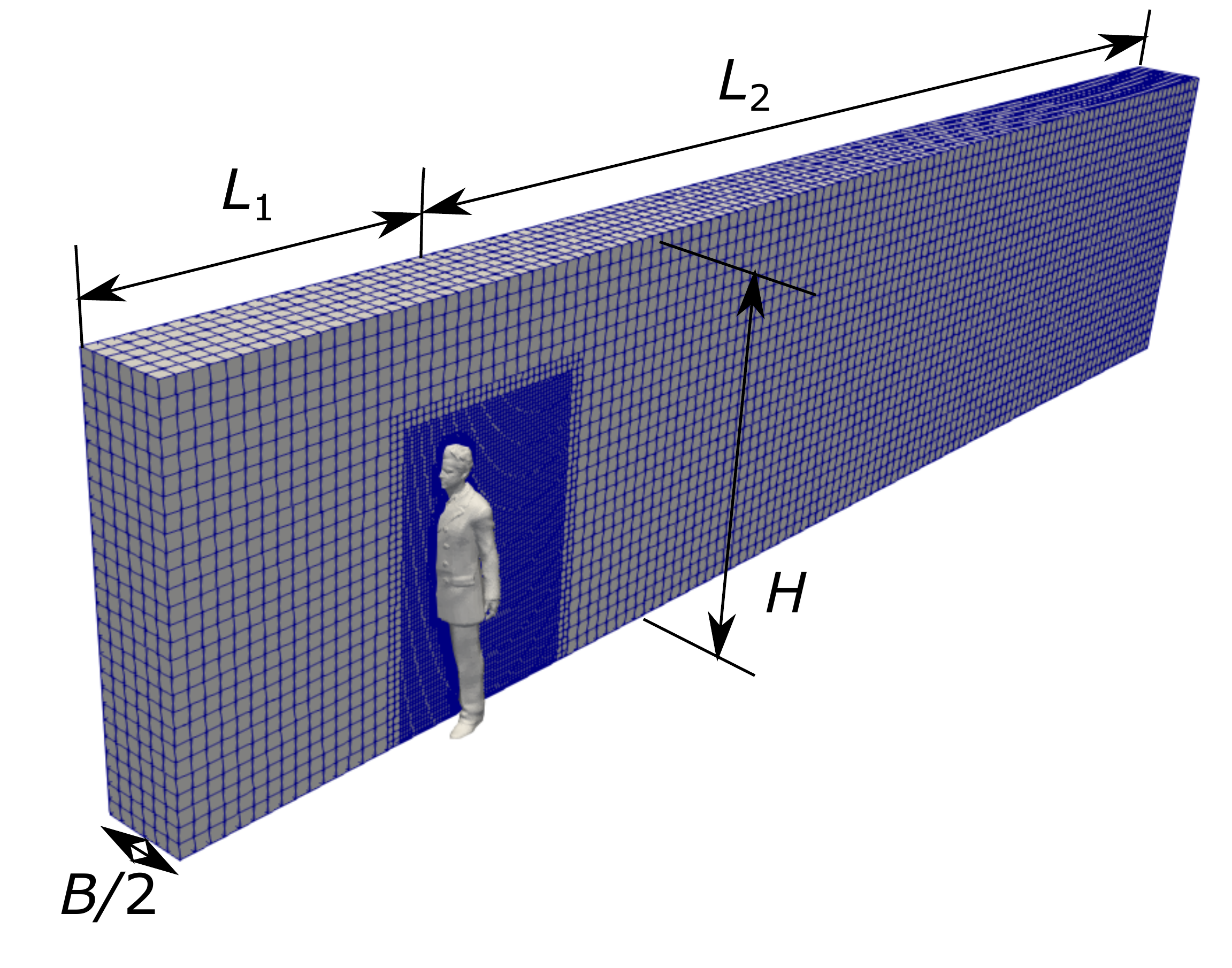}
	\caption{Computational domain and mesh configuration.}
	\label{fig:computationalDomain}
\end{figure}

The computational domain is rectangular with a central symmetric plane (see Fig. \ref{fig:computationalDomain}). The domain height is $H = 2.8$~m, the length in front of the manikin is $L_1 = 2.0$~m, and the length behind the manikin is $L_2 = 10.0$~m. Different domain breadths from $B = 1.2$ m to $B =6.0$ m are considered to analyze the influence of space sizes. 
In the flow simulation, the reference frame is fixed to the manikin with air blowing from the inlet. Free-slip condition is imposed on the lateral, top, and bottom boundaries.  Non-slip condition is imposed on the manikin with wall function \cite{SpaldingWall}. 
The simulations represent a daily scenario where a man walks at a constant speed in the range of $U \in [1.2, 1.8]$ m/s and coughs. The Reynolds number based on the shoulder breadth and the walking speed is in the range of $Re \in \left[ 3.6 \times 10^4, 5.4 \times 10^4 \right]$. 

In this work, the droplets are modelled as a cloud of spheroid particles and are injected into the computational domain from the mouth of the manikin within 0.12 s at the beginning.  The droplet diameter distribution follows the Weibull distribution. The diameter range is $d \in [1.0, 300.0]$ $\mu$m and the mean diameter is 80.0 $\mu$m.  All particles are emitted with a horizontal velocity of $v_c^x = 5.0$ m/s. These characteristics follow a recent CFD analysis of cough droplet dispersion \cite{droplets}.  In the simulation of the droplet dispersion, the ground fixed reference frame is employed and the manikin moves from the right to the left.

\begin{figure*}
	\centering
		\includegraphics[width=.45\textwidth]{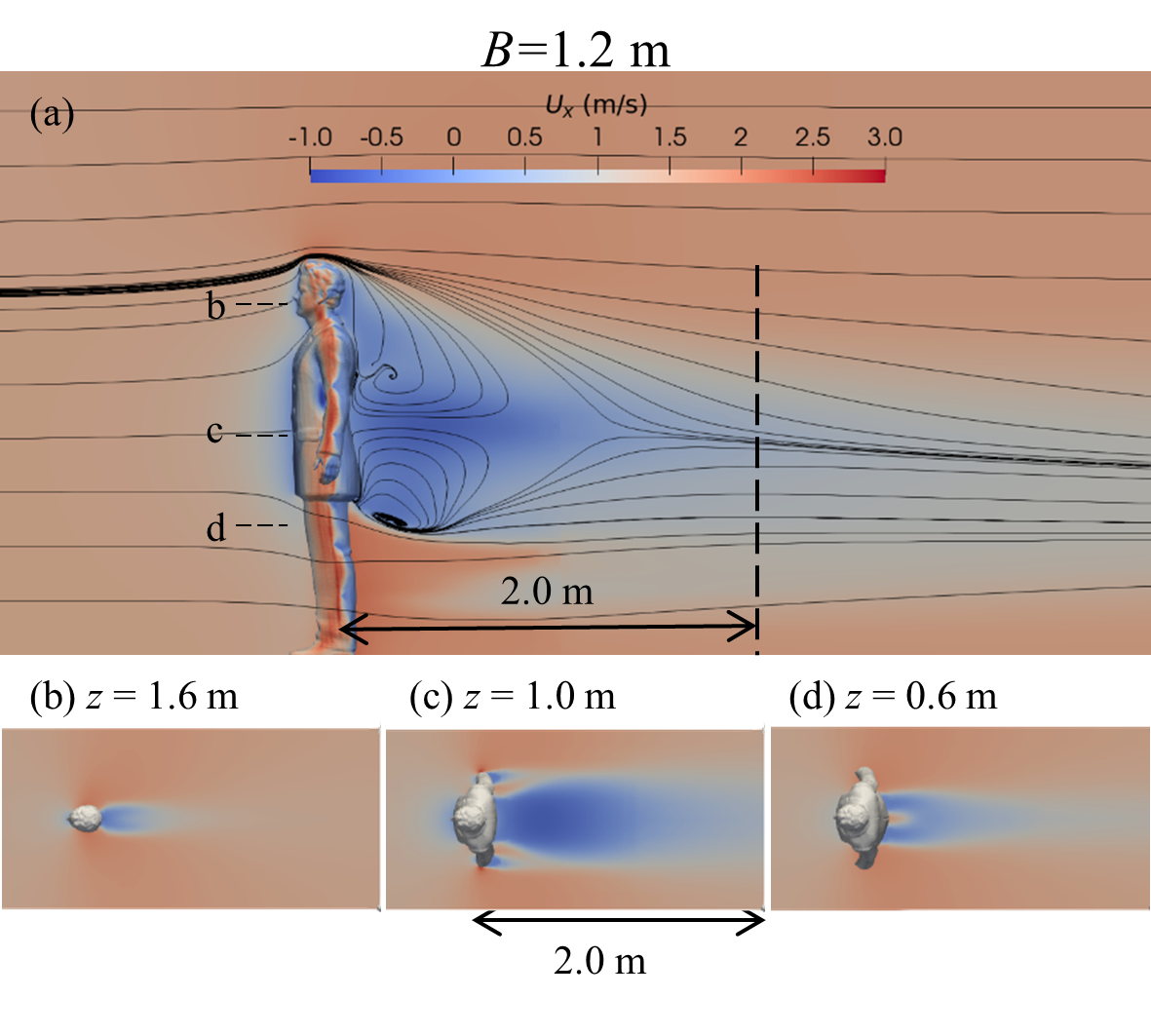}
		\includegraphics[width=.45\textwidth]{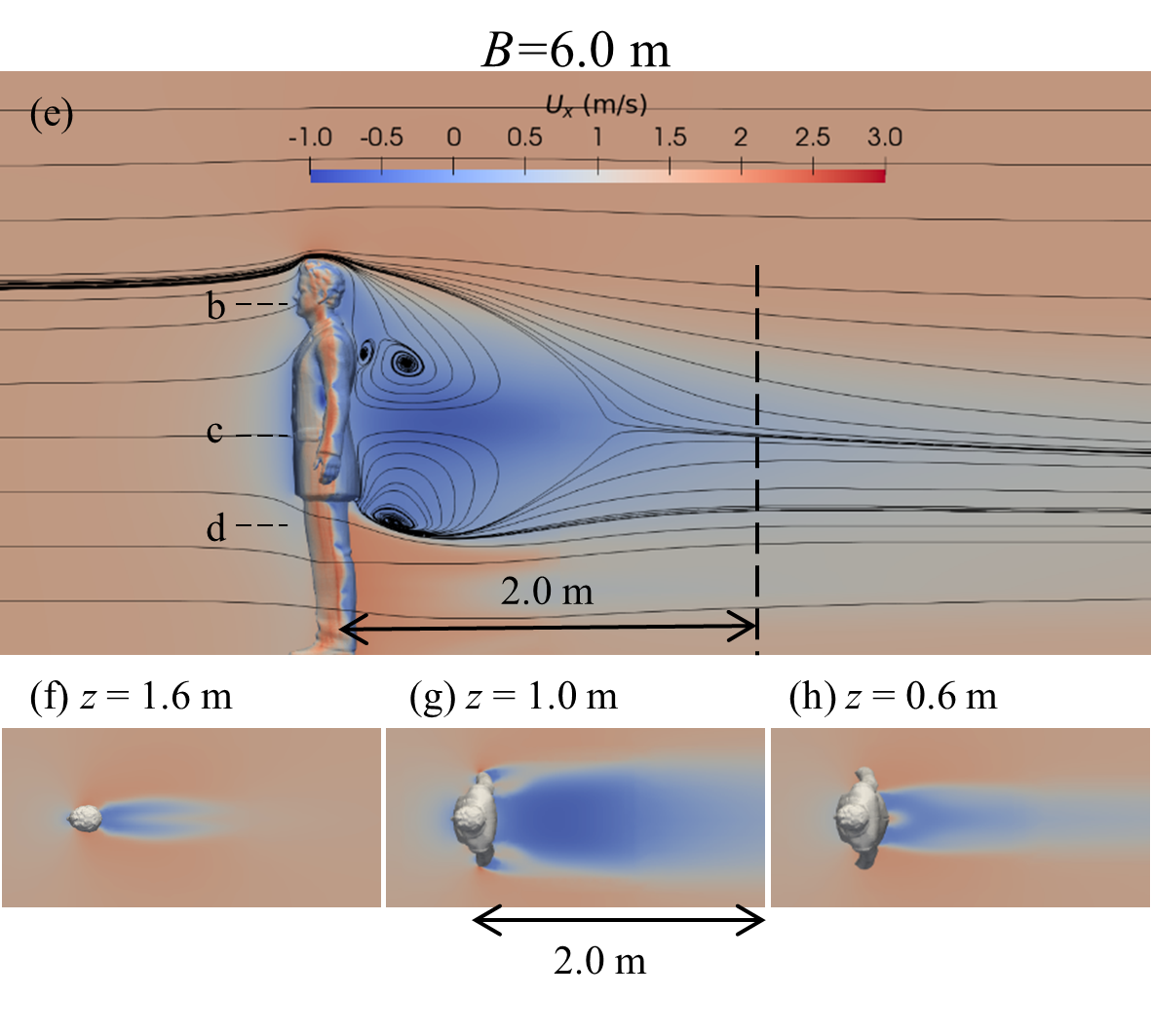}
	\caption{Patterns of air flow around the manikin for the cases with $B=1.2$ m (left) and $B=6.0$ m (right) and walking speed 1.5 m/s. Figures on the top: contours of streamwise velocity and streamlines on the symmetrical plane; figures on the bottom: contours of streamwise velocity on horizontal planes located at different vertical locations.}
	\label{fig:flowfield}
\end{figure*}

Figure \ref{fig:flowfield} shows the flow field around the manikin for $B=1.2$ m and $6.0$ m with walking speed $U = 1.5$ m/s. In Figs. \ref{fig:flowfield} (a) and (e), the vertical dashed lines are two meters away from the manikin. The small panels represent the flow field on horizontal planes and are clipped at two meters downstream from the manikin, to indicate the commonly practiced social-distancing. Panels on the left are for the case with $B=1.2$ m, and those on the right are for the case with $B=6.0$ m, respectively. In general, the manikin decelerates the flow around it as bluff-body. But the flow patterns are found being strongly related to the shape of the  human body. As seen in Fig. \ref{fig:flowfield} (a), a reverse flow region exists behind the head and the torso with the maximum reverse velocity locating approximately at the waist-height ($z = 1.0$~m, marked by $c$). Behind the legs, the flow is slightly faster than the ambient flow. Figures \ref{fig:flowfield} (b) (c) and (d) show the wake behind the mouth, the waist, and the legs. As seen, the manikin's torso part (Fig. \ref{fig:flowfield} (c)) induces the strongest wake. Special wake patterns, such as the jets through the gaps between the hands and the torso (in Fig. \ref{fig:flowfield} (c)) and the gap between the legs (in Fig. \ref{fig:flowfield} (d)) are also remarkable, in which the latter one explains the high speed flow below the waist shown in Fig. \ref{fig:flowfield} (a). At two meters downstream, the wake is almost negligible at the mouth height and the leg height as shown in Fig. \ref{fig:flowfield} (b) and (d), respectively,  while it is still visible at the waist height as shown in Fig. \ref{fig:flowfield} (c). For the case with $B = 6.0$ m, similar flow patterns are observed in Figs. \ref{fig:flowfield} (e) (f) (g) and (h), but with a slightly larger re-circulation region behind the torso and a slightly stronger velocity deficit as compared with the case with $B=1.2$ m.  

\begin{figure}
	\centering
		\includegraphics[width=0.45\textwidth]{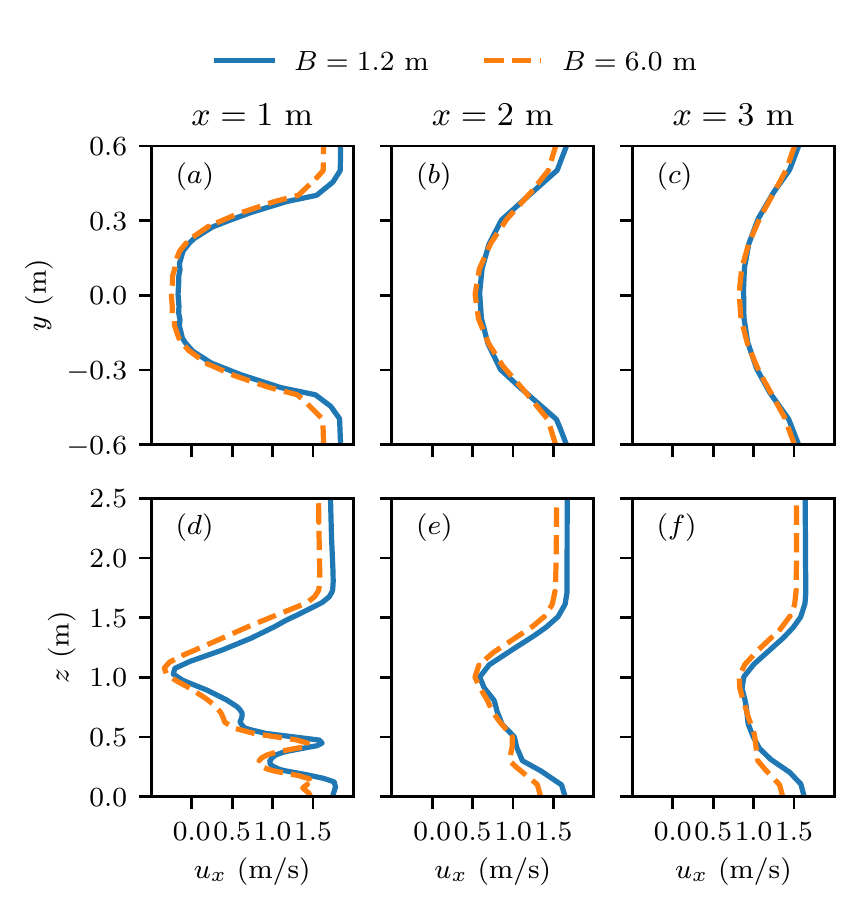}
	\caption{Comparison of streamwise velocity profile behind the manikin for the cases with $B=1.2$ and $6.0$ m and manikin's walking speed $1.5$ m/s with (a) (b) (c) for  transverse profiles of the streamwise velocity $u_x$ at the waist height $z=1.0$~m, and (d) (e) (f)  for vertical profiles of $u_x$ on the symmetry plane.}
	\label{fig:flowCompare}
\end{figure}

Figures \ref{fig:flowCompare} (a) (b) and (c) compare the transverse profiles of the streamwise velocity in the wake between the cases with $B=1.2$ m and $6.0$ m at $U = 1.5 $ m/s on the waist height. As seen, the differences manifest mostly in the near wake. At one meter behind the manikin (Fig. \ref{fig:flowCompare} (a)), the major difference exists in the velocity overshoot region outside the wake, where the streamwise velocity of the case with $B=1.2$ m is approximately 10\% larger than that of the case with $B=6.0$ m.  In Fig. \ref{fig:flowCompare} (d), the vertical profiles of the streamwise velocity are compared. As seen, the two vertical profiles are similar with each other with complex variations and the largest velocity deficit located around the waist height. In both cases, the upper boundary of the wake region is approximately at 1.5 m above the ground at $x = 1$ m downstream of the manikin and gradually decreases at further downstream locations.  Outside the wake in the vertical direction, the streamwise velocity is also higher for the case with $B=1.2$ m because of the higher blockage effect due to the narrower space.  
Traveling downstream, the wake recovers with momentum exchange with the free stream in both cases, and the differences between the two cases become smaller, especially for the profiles in the transverse direction as shown in Fig. \ref{fig:flowCompare} (c).

\begin{figure*}
	\centering
		\includegraphics[width=0.45\textwidth]{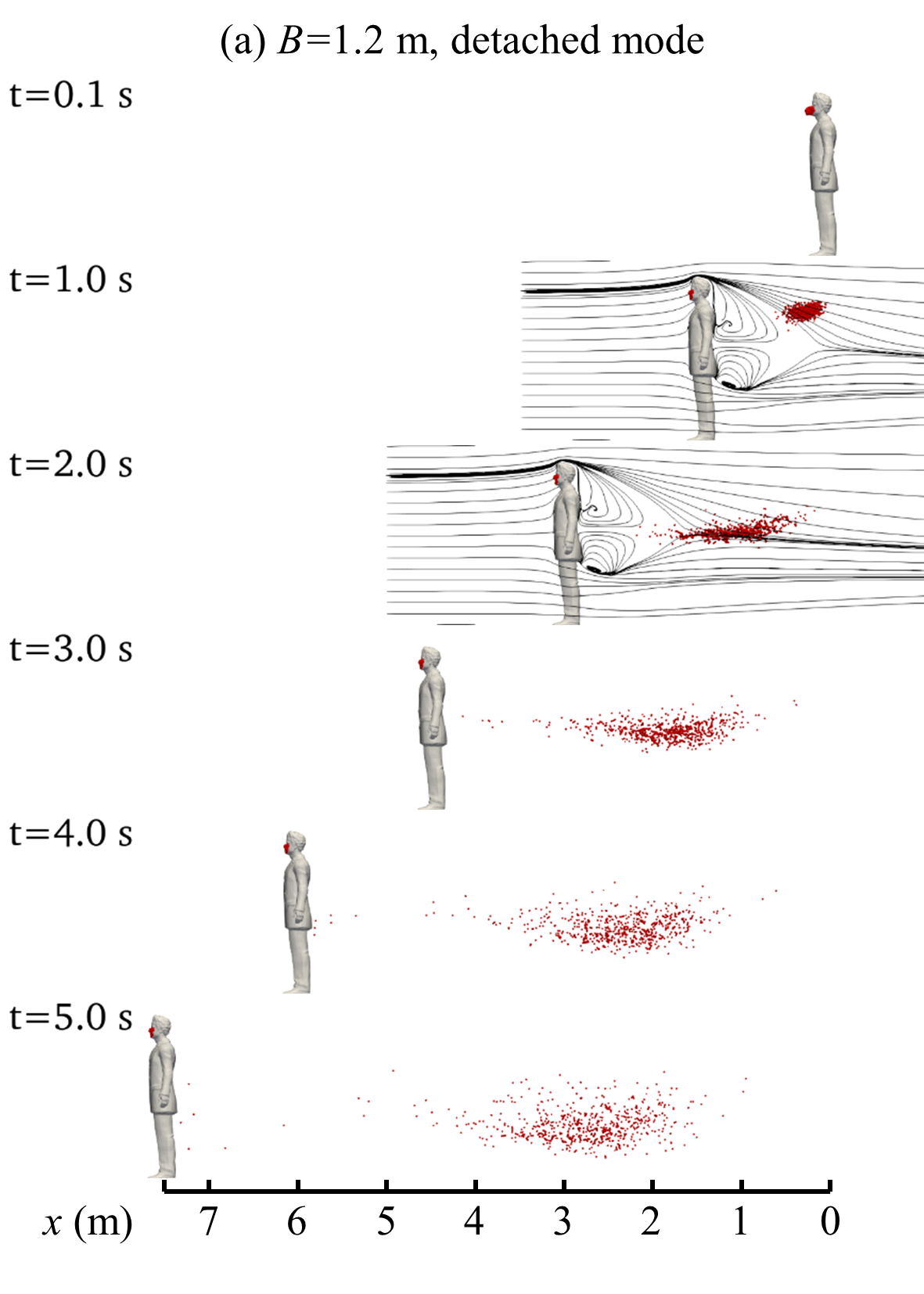}
		\hspace{1cm}
		\includegraphics[width=0.45\textwidth]{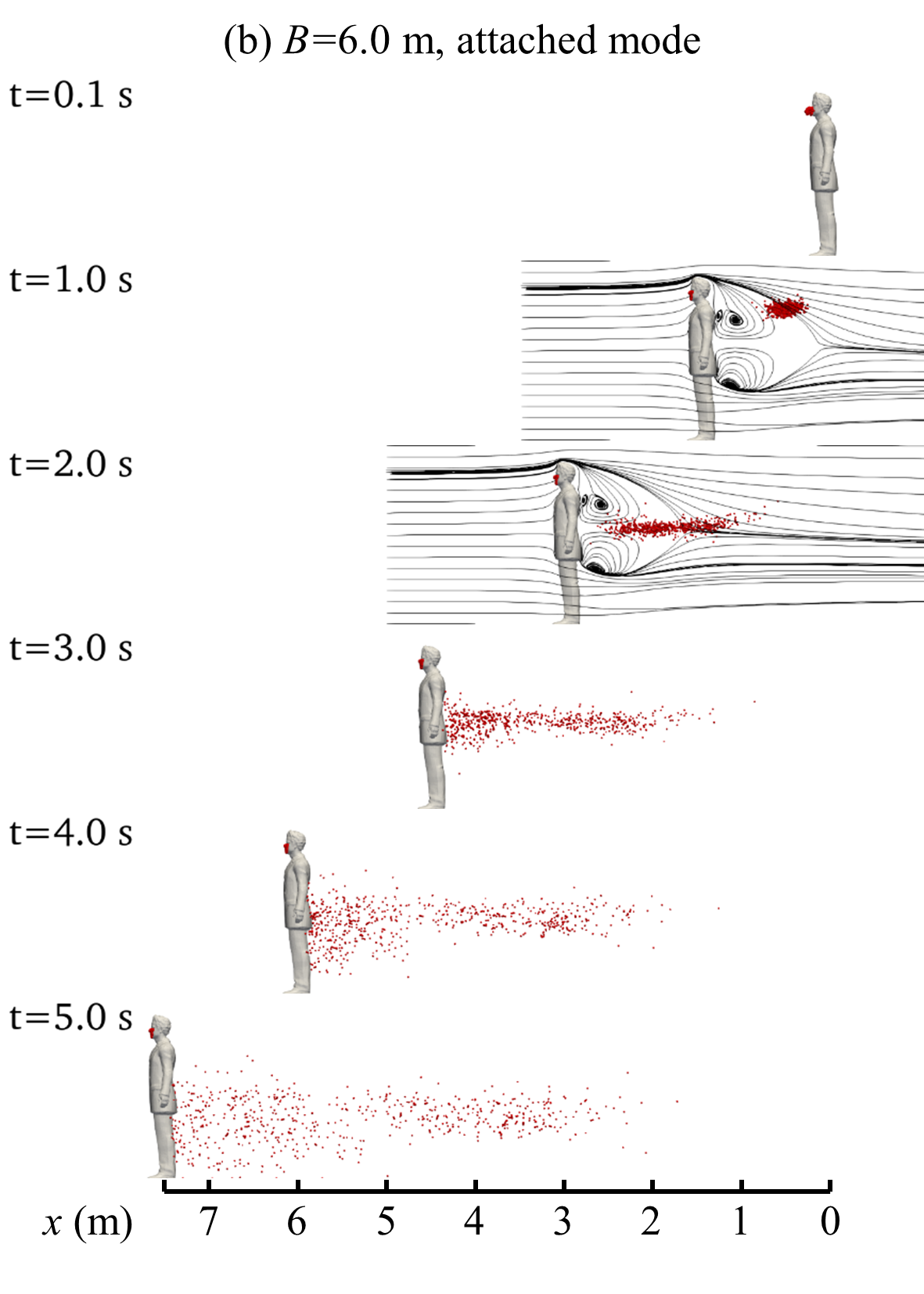}
	\caption{Patterns of droplet dispersion in the wake of the walking manikin for : (a) $B=1.2$ m case, (b) $B=6.0$ m case,  with walking speed 1.5 m/s. Droplets are plotted as red dots of the same size regardless of their real diameter. At $t =  1.0$ s and $2.0$ s, the black streamlines are added to illustrate the range of the  re-circulation bubbles.}
	\label{fig:particle}
\end{figure*}

After showing the flow field, we examine the dispersion of droplets in Fig. \ref{fig:particle} for five different instants. At $t = 0.1$ s, a small cloud of droplets, marked as red dots, releases from the manikin's mouth. At $t = 1.0$ s, the cloud of droplets expands in size and advances at a velocity smaller than the manikin. For both cases, the cloud is of  oval shape at $t = 1.0$ s from the side view, which is similar to the simulation result of cough droplets carried by mild wind without wake effect in the work of Dbouk and  Drikakis \cite{droplets}. This similarity of the shape of the droplet cloud between the present work and that in the reference, in which the head is not considered, implies that the head-induced wake has very limited effects on the droplet motion. The oval shape deformation of the droplet cloud can be explained as the result of different advancing and settling velocities of droplets of different sizes \cite{wang_prf}. At $t = 2.0$ s, the cloud drops approximately to the waist height and deforms into an elongated shape that expands horizontally for both cases. Significant differences between the two cases are observed starting from this instant. For the case with $B=1.2$ m, the cloud is left further behind the manikin, and at $t = 5.0$ s the cloud locates in the range of $x \in \left[ 2,4\right]$ m, leaving the region just behind the manikin ($x \in [4, 7.5]$ m) nearly unaffected. In the case with $B=6.0$ m, on the other hand, a part of the droplet cloud moves towards the manikin from $t = 1.0$ s to $t = 2.0$ s. From $t = 3.0$ to 5.0 s, the left limit of the cloud catches up with the manikin and the droplet cloud extends in a region much broader than the case with $B=1.2$ m. 

\begin{figure}
	\centering
		\includegraphics[width=0.45\textwidth]{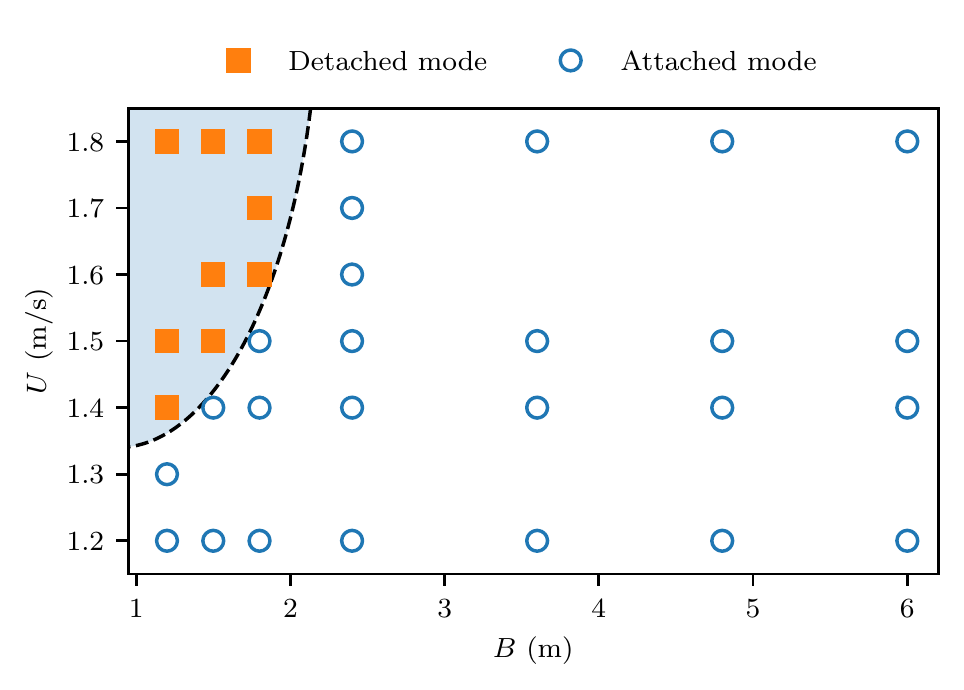}
	\caption{The mode map in the two-dimensional space of the domain breadth $B$ and the walking speed $U$. }
	\label{fig:modeMap}
\end{figure}

These two significantly different patterns are referred to as the attached mode and the detached mode hereafter. In the case with $B = 6.0$ m, the attached mode is formed because a large portion of the droplets falls into the re-circulation bubble as shown by Fig. \ref{fig:particle} (b) in the snapshots at $t = 1.0$ s and 2.0 s. In the case with $B=1.2$ m, on the other hand, the slightly larger overshoot velocity above the wake and the smaller re-circulation bubble help the droplets escape from the reverse flow region, forming the  detached mode.  A mode map for different space sizes (different $B$ values) and different walking speed ($U$) is shown in Fig. \ref{fig:modeMap}. As seen, the  attached mode  exists for most cases, while the detached mode  is only observed for cases with narrower space and higher walking speed.


In summary, we investigated the effects of space sizes on the dispersion of cough droplets behind a walking manikin of realistic shape using the RANS method. Similar flow patterns are observed for cases with different space constraints. For all the considered cases, suspension of droplets  is observed below the waist height of the manikin for distance larger than two meters, indicating higher risks for kids who walk behind an coughing patient when following the current social-distancing guideline. More importantly, two distinct particle dispersion modes, \textit{i.e.}, the  attached mode  and the  detached mode,   are discovered  for different space sizes. The detached mode only occurs for cases with small space and high walking speed, while the attached mode happens for other cases. When the attached mode happens, the cloud of droplets is observed starting from the rear of the manikin with elongated shape in the streamwise direction. For the detached mode, on the other hand, the cloud of droplets is separated from the manikin and convected at much lower speed, with its size in the streamwise direction much smaller and the droplet concentration remarkably higher than that of the attached mode at five seconds after coughing. This poses a great challenge on determining the safe distance for places with high space constraint,  \textit{e.g.}, in a very narrow corridor,  as a person may still inhale viral droplets even the patient is far in front of him/her.  


In this work, only the steady-state part of the air flow is explicitly taken into account for the dispersion of droplets without considering the flow unsteadiness, and with the effect of turbulent fluctuations approximated with a kinematic model. Methods of higher fidelity, such as Large-Eddy Simulation or Direct Numerical Simulation \cite{zhou2018structural}, with the capacity to predict  the unsteady, turbulent motion of the wake flow and the droplet dispersion,  could be employed in the future to study the present cases in more detail.      

This work is partially supported by NSFC Basic Science Center Program for ``Multiscale Problems in Nonlinear Mechanics" (NO. 11988102). 

The data that support the findings of this study are available from the corresponding author upon reasonable request.

\bibliography{main}

\end{document}